\def\ra{\rangle}
\def\la{\langle}
\def\be{\begin{equation}}
\def\ee{\end{equation}}
\def\ba{\begin{array}}
\def\ea{\end{array}}

\documentclass[amsfonts,showpacs,twocolumn,pra]{revtex4}
\usepackage{graphicx}
\usepackage{epstopdf}
\usepackage{epsfig}
\usepackage{subfigure}
\usepackage{caption2}
\usepackage{amsmath}
\usepackage{amssymb}
\def\qed{\leavevmode\unskip\penalty9999\hbox{}
         \nobreak\hfill
     \quad\hbox{\leavevmode  \hbox to.77778em{\hfil\vrule  \vbox to.675em {\hrule width.3em\vfil\hrule}\vrule\hfil}}
     \par\vskip 0pt}

\begin{document}
\setcounter{secnumdepth}{3} % ±íʾsection²ã´ÎÉî¶ÈΪ3
\renewcommand\thesection{\Roman{section}}
\renewcommand\thesubsection{\Alph{subsection}}

\title{Quantifying Algebraic Asymmetry of Hamiltonian Systems}
\author{Hui-Hui Qin$^{1}$}
\author{Shao-Ming Fei$^{2,3}$}
\author{Chang-Pu Sun$^{1,4}$}
\affiliation{$^1$ Beijing Computational Science Research Center, Beijing 100193, China\\
$^2$ School of Mathematical Sciences, Capital Normal University,
Beijing 100048, China\\
$^3$ Max-Planck-Institute for Mathematics in the Sciences, Leipzig 04103, Germany\\
$^4$ Graduate School of China Academy of Engineering Physics, Beijing 100193, China}

\begin{abstract}
The symmetries play important roles in physical systems. We study the symmetries of a Hamiltonian system by investigating the asymmetry of the Hamiltonian with respect to certain algebras.
We define the asymmetry of an operator with respect to an algebraic basis in terms of their commutators. Detailed analysis is given to the Lie algebra $\mathfrak{su}(2)$ and its $q$-deformation. The asymmetry of the $q$-deformed integrable spin chain models is calculated.
The corresponding geometrical pictures with respect to such asymmetry is presented.
\end{abstract}
\pacs{03.67.Mn, 03.67.-a, 02.20.Hj, 03.65.-w}
\maketitle

\section{Introduction}

The symmetry of a physical system plays a central role in a broad range of problems in physics \cite{Wigner,Wigners,Dirac}. In N\"{o}ther¡¯s theorem \cite{Noether} symmetries are used to find the corresponding conservation laws of dynamics. The symmetries of a Hamiltonian also help to obtain the solutions of the evolution equations, as well as the degeneracy of the energy levels of the physical systems \cite{Wigners,Weinberg}. Some important internal symmetries such as the identity principle of micro-particles have been also revealed \cite{Landau}. In studying the integrable models the theory of Lie groups and Lie algebras \cite{Wigners,Weyl,Hamermesh}, as well as the quantum groups \cite{Kulish,Faddeev,Chaichian} and quantum algebras \cite{Macfarlane,Biedenharn,Sun} has been effectively used.

The quantification of the symmetry or asymmetry of a quantum system is of significance. The asymmetry of quantum states with respect to groups has been studied in the view of resource, such as coherence \cite{Gour,Vaccaro,Toloui,Skotiniotis,Iman,Piani,Luo}. In \cite{Wakakuwa} the cost of symmetrization of quantum states has been investigated. In \cite{Fang,Dong,Dong2} the authors considered the asymmetry measure of a Hamiltonian with respect to Lie group transformations, aiming at revealing some phenomena such as symmetry and spontaneous symmetry breaking, and quantum phase transition caused by the symmetry changes of Hamiltonian systems.

It is said that an operator $H$ is symmetric with respect to a given algebra $\mathfrak{g}$, if it commutates with all the generators of $\mathfrak{g}$. An example of such $H$ is the Casimir operator of $\mathfrak{g}$. However, if $H$ does not commutate with all the generators of $\mathfrak{g}$, one needs to assure the degree of asymmetry of $H$ with respect to $\mathfrak{g}$. Geometrically, one may ask what the degree of asymmetry of a deformed sphere, say ellipsoid, with respect to the standard sphere is.

In this paper we study the asymmetry of an operator with respect to algebras including the usual Lie algebras, Heisenberg algebras \cite{Heisenberg}, Lorentz algebras \cite{Schmidke} and Hopf algebras \cite{Hopf} etc.. We give a definition of the asymmetry degree, and analyze in detail the systems with $SU(2)$ symmetry and the deformed $SU_q(2)$ symmetry. The harmonic oscillator systems and the integrable spin chain systems are also taken into account. The explicit geometrical pictures associated with the asymmetries are presented.

\section{Asymmetry of operators with respect to an algebraic basis}

Denote $\{X_j\}$ the basis of an algebra $\mathfrak{g}$.
The symmetry of an operator $H$ with respect to $\mathfrak{g}$ is judged by the commutation relations between $H$ and the basis $\{X_j\}$: $[H,X_j]=0$, $\forall j$. If there exists $X_j$ such that $[H,X_j]\neq 0$, $H$ is not symmetric with respect to $\mathfrak{g}$. We quantify the non-commutative extent between $H$ and $X_j$ to characterize the degree of asymmetry of $H$ with respect to $\mathfrak{g}$. Denote $\|O\|^2=tr(O^{\dagger}O)$ the Frobenius norm of operator $O$. We define the asymmetry degree of the operator $H$ with respect to $\mathfrak{g}$ as follows,
\be\label{def1}
\mathcal{A}(\mathfrak{g},H)=
\frac{1}{\|\tilde{H}\|^2}\sum_{j} \|[H,X_j]\|^2,
\ee
where $\tilde{H}=H-\frac{tr(H)}{d}I$, $d$ is the order of the operator $H$ and $I$ is the identity operator of $d$ orders.

From the definition (\ref{def1}), $\mathcal{A}(\mathfrak{g},H)$ has the following properties:

(\uppercase\expandafter{\romannumeral1}). $\mathcal{A}(\mathfrak{g},H)\geq 0$, the equality holds if and only if $[H,X_j]=0$, $\forall j$. Namely, $\mathcal{A}(\mathfrak{g},H)=0$ if and only if $H$ commutates with $X_j$ for all $j$. We call $H$ symmetric with respect to $\mathfrak{g}$, or $H$ is $\mathfrak{g}$-symmetric in this case.

(\uppercase\expandafter{\romannumeral2}). $\mathcal{A}(\mathfrak{g},H)\geq \mathcal{A}(\mathfrak{g},THT^{-1})$, where $T$ is an invertible transformation which commutes with $\mathfrak{g}$, $[T,X_j]=0$ $\forall j$. This property can be seen as following,
\be
\begin{aligned}
\mathcal{A}&(\mathfrak{g},THT^{-1})\\
&=\sum_j tr([X_j,THT^{-1}]^{\dagger}[X_j,THT^{-1}])\\
&=\sum_j tr([X_j,H]^{\dagger}(T^{\dagger}T)
[X_j,H](T^{\dagger}T)^{-1})\\
&\leq \sum_j tr([X_j,H]^{\dagger}[X_j,H])
=\mathcal{A}(\mathfrak{g},H).
\end{aligned}
\ee
Therefore, the asymmetry degree of $H$ is monotonically non-increasing under $\mathfrak{g}$-symmetric transformations.

(\uppercase\expandafter{\romannumeral3}). $\mathcal{A}(\mathfrak{g},H+\lambda I)=\mathcal{A}(\mathfrak{g},H)$ for arbitrary parameter $\lambda\in \mathbb{R}$, which implies that $\mathcal{A}(\mathfrak{g},H)$ is independent of constant potential energy, i.e., the asymmetry keeps invariant under shifting the ground energy of the Hamiltonian $H$ of the system.

We now compute several examples in detail.

\subsection{The harmonic and deformed harmonic systems}

For a two linear oscillator system, the Hamiltonian has the form,
\be\label{suh}
H=a^{\dagger}_1a_1+a^{\dagger}_2a_2,
\ee
where $a_j$ and $a^{\dagger}_j$, $j=1,2$, are the creation and annihilation operators \cite{Dirac},
satisfying the commutation relations $[a_j,a^{\dagger}_k]=\delta_{jk}I$, $j,k=1,2$.

Underlying the Hamiltonian system (\ref{suh}) is the $\mathfrak{su}(2)$ algebra, with the generators $J_1$, $J_2$ and $J_3$ satisfying
\be\label{su}
[J_i,J_j]=\epsilon_{ijk} J_k,~~~i,j,k=1,2,3.
\ee
where $\epsilon_{ijk}$ is the Levi-Civita symbol. In terms of the creation and annihilation operators, $J_1$, $J_2$ and $J_3$ have the following explicit representations,
\be\label{hsu2}
\begin{aligned}
&J_+=J_1+i J_2=a^{\dagger}_1a_2, \qquad J_-=J_1-i J_2=a^{\dagger}_2a_1, \\
&J_3=\frac{1}{2}(a^{\dagger}_1a_1 -a^{\dagger}_2a_2)=\frac{1}{2}(N_1-N_2),
\end{aligned}
\ee
where $N_j=a^{\dagger}_ja_j$ is the number operator defined by the creation and annihilation operators $a^{\dagger}_j$ and $a_j$, $(j=1,2)$.
It is easily verified that the system $H$ has $\mathfrak{su}(2)$ symmetry, $[H,J_i]=0$ for $i=1,2,3$.

In addition to the usual Lie algebras, one also has the universal enveloping algebras, the quantum (q-deformed) Lie algebras \cite{Jimbo1,Drinfeld}. With respect to the Lie algebra $\mathfrak{su}(2)$, the quantum algebra $\mathfrak{su}_q(2)$ is given by the generators
$J'_\pm=J'_1\pm i J'_2$ and $J'_3=J_3$, satisfying the following algebraic relations,
\be\label{suq}
[J_3,J'_{\pm}]=\pm J'_{\pm},~~[J'_+,J'_-]=[2J_3]_q,
\ee
where $[O]_q=\frac{q^{O}-q^{-O}}{q-q^{-1}}$.

Correspondingly, one has the $q$-deformed quantum harmonic oscillators \cite{Chaichian} and the
irreducible representation of $\mathfrak{su}_q(2)$ \cite{Macfarlane,Biedenharn,Sun,Chang,Fei}.
The $q$-deformed quantum harmonic oscillators are given by $b^{\dagger}_j$ and $b_j$ satisfying
\be\label{qh}
b_jb^{\dagger}_k-q^{-1}b^{\dagger}_j b_k=q^{N_j}\delta_{jk},~~j,k=1,2,
\ee
where $N_j=a^{\dagger}_ja_j$, $j=1,2$.

From (\ref{qh}) one has the representations,
\be\begin{aligned}
&J'_+=b^{\dagger}_1b_2, ~~~ J'_-=b^{\dagger}_2b_1,\\
&J'_3=J_3=\frac{1}{2}(N_1-N_2).
\end{aligned}
\ee

It is straightforward to verify that the $q$-deformed Hamiltonian $H'$,
\be\label{hp}
H'=b^{\dagger}_1b_1+b^{\dagger}_2b_2,
\ee
is $\mathfrak{su}_q(2)$-symmetric, i.e., $[H',J'_i]=0$ for $i=1,2,3$.

However, the q-deformed harmonic Hamiltonian $H'$ is no longer $\mathfrak{su}(2)$ symmetric.
In order to calculate the asymmetry degree of $H'$ with respect to $\mathfrak{su}(2)$,
consider the Fock representations of $\{a^{\dagger}_j,a_j,N_j\}$ and $\{b^{\dagger}_j,b_j,N_j\}$ \cite{Macfarlane,Biedenharn,Sun}, respectively,
$$
\begin{aligned}
&|m\ra=|m_1,m_2\ra=\Big(
\frac{a^{\dagger m_1}_1a^{\dagger m_2}_2}{\sqrt{m_1!}\sqrt{m_2!}}\Big)|0\ra,\\
&|m\ra'=|m_1,m_2\ra'=\Big(
\frac{b^{\dagger m_1}_1b^{\dagger m_2}_2}{\sqrt{[m_1]_q!}\sqrt{[m_2]_q!}}
\Big)|0\ra,
\end{aligned}
$$
where the vacuum state $|0\ra$ satisfies $a_j|0\ra=b_j|0\ra=0$, $j=1,2$, and
$[m]_q!=[m]_q[m-1]_q\cdots[2]_q[1]_q$.

The generators of $\mathfrak{su}(2)$ have the following representations,
\be
\begin{aligned}
&J_+|m\ra=a^{\dagger}_1a_2|m\ra=
\sqrt{(m_1+1)m_2}|m_1+1,m_2-1\ra;\\
&J_-|m\ra=a^{\dagger}_2a_1|m\ra=
\sqrt{(m_2+1)m_1}|m_1-1,m_2+1\ra;\\
&J_3|m\ra=\frac{1}{2}(N_1-N_2)|m\ra
=\frac{1}{2}(m_1-m_2)|m_1,m_2\ra.
\end{aligned}
\ee
Since $N_j|m\ra=m_j|m\ra$ and $N_j|m\ra'=m_j|m\ra'$, the Hamiltonian $H'$ can be also represented as
$$
H'|m\ra=(b^{\dagger}_1b_1+b^{\dagger}_2b_2)|m\ra=([m_1]_q+[m_2]_q)|m_1,m_2\ra.
$$
On a finite dimensional subspace with basis $\{|m\ra=|m_1,m_2\ra|m_1+m_2=M|1<M<\infty\}$, we have that
\begin{widetext}
\be
\begin{aligned}
&[H',J_+]|m\ra=
\left\{
\begin{aligned}
&\sqrt{(m_1+1)m_2}
([m_1+1]_q+[m_2-1]_q-[m_1]_q-[m_2]_q)|m_1+1,m_2-1\ra,m_2>0\\
&0,m_2=0,
\end{aligned}
\right.\\
&[H',J_-]|m\ra=
\left\{
\begin{aligned}
&\sqrt{m_1(m_2+1)}
([m_1-1]_q+[m_2+1]_q-[m_1]_q-[m_2]_q)|m_1-1,m_2+1\ra,m_1>0\\
&0,m_1=0,
\end{aligned}
\right.\\
&[H',J_3]|m\ra=0.
\end{aligned}
\ee
\end{widetext}
Then the asymmetry degree of $H'$ with respect to $\mathfrak{su}(2)$ is then given by,
\begin{widetext}
\be\label{eqasymmetryH1}
\begin{aligned}
\mathcal{A}(\mathfrak{su}(2),H')
&=\frac{1}{\|\tilde{H'}\|^2}
\sum_{\alpha=\pm,3}\|[H',J_{\alpha}]\|^2
=\frac{1}{R}\sum_{m_1+m_2=M}\sum_{\alpha=\pm,3}
\la m_1,m_2|[H',J_{\alpha}]^{\dagger}
[H',J_{\alpha}]|m_1,m_2\ra\\
&=\frac{2}{R}\sum_{m_1+m_2=M}
(m_1+1)m_2([m+1]_q+[m_2-1]_q-[m_1]_q-[m_2]_q)^2\\
&=\frac{2}{R}\sum^{M}_{j=1}j(M+1-j)
\frac{(\cosh\gamma(M-j+\frac{1}{2})-
\cosh\gamma(j-\frac{1}{2}))^2}
{\cosh^2\gamma},
\end{aligned}
\ee
\end{widetext}
where $R=\|\tilde{H'}\|^2=\sum^{M}_{j=1}([j]_q+[M-j]_q)^2-
\frac{4(\sum^{M}_{j=1}[j]_q)^2}{M+1}$ and $\gamma=\log q$.

\begin{figure}[!htb]
  \centering
  \includegraphics[width=0.6\columnwidth]
  {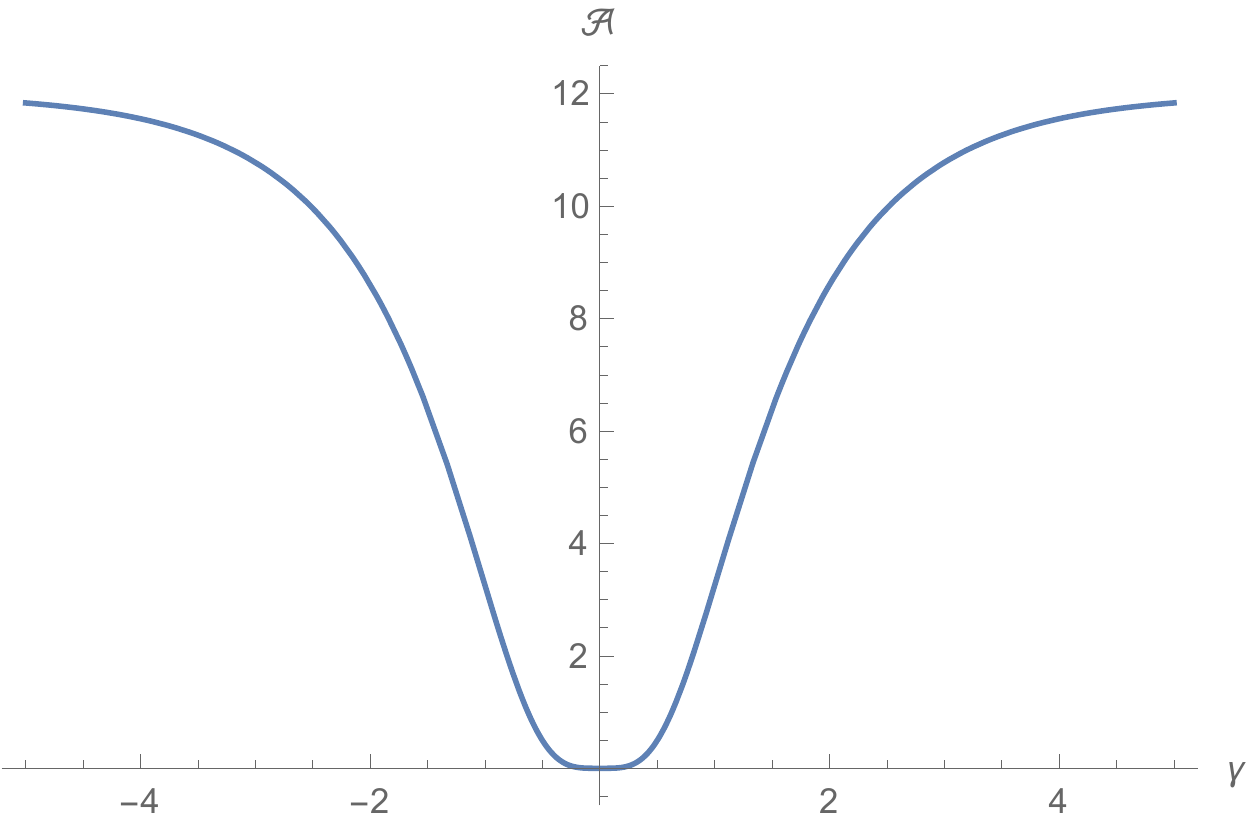}\\
  \caption{The asymmetry degree of the operator $H'$ with respect to the algebra $\mathfrak{su}(2)$, as a function of the deformation parameter $\gamma$.}\label{figplus}
\end{figure}

Fig. \ref{figplus} shows the asymmetry of the operator $H'$ with respect to the algebra $\mathfrak{su}(2)$ for $M=2$.
From (\ref{eqasymmetryH1}) and Fig. \ref{figplus} it is obvious that when $q\rightarrow 1$ ($\gamma\rightarrow 0$), $\mathcal{A}(\mathfrak{su}(2),H')
\rightarrow0$, i.e., $H'\rightarrow H$ is $\mathfrak{su}(2)$ symmetric.

\subsection{The geometrical characterization of asymmetry: $SU(2)$ v.s. deformed $SU_q(2)$}

\begin{figure}[!htb]
\centering
\subfigure[$q=1(\gamma=0)$]{
\label{Fig.sub.1}
\includegraphics[width=0.4\columnwidth]
{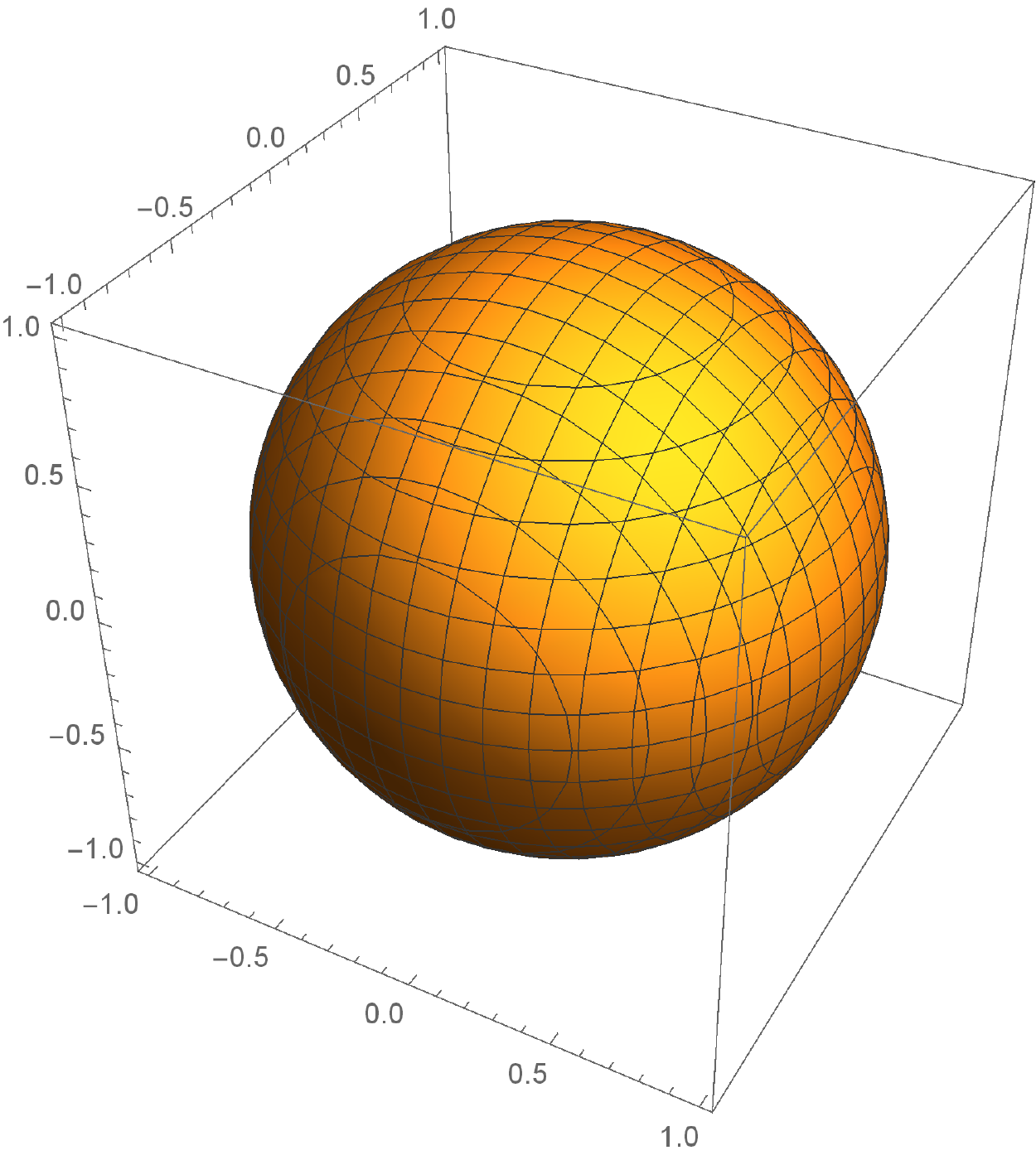}}
\subfigure[$q=e(\gamma=1)$]{
\label{Fig.sub.2}
\includegraphics[width=0.4\columnwidth]
{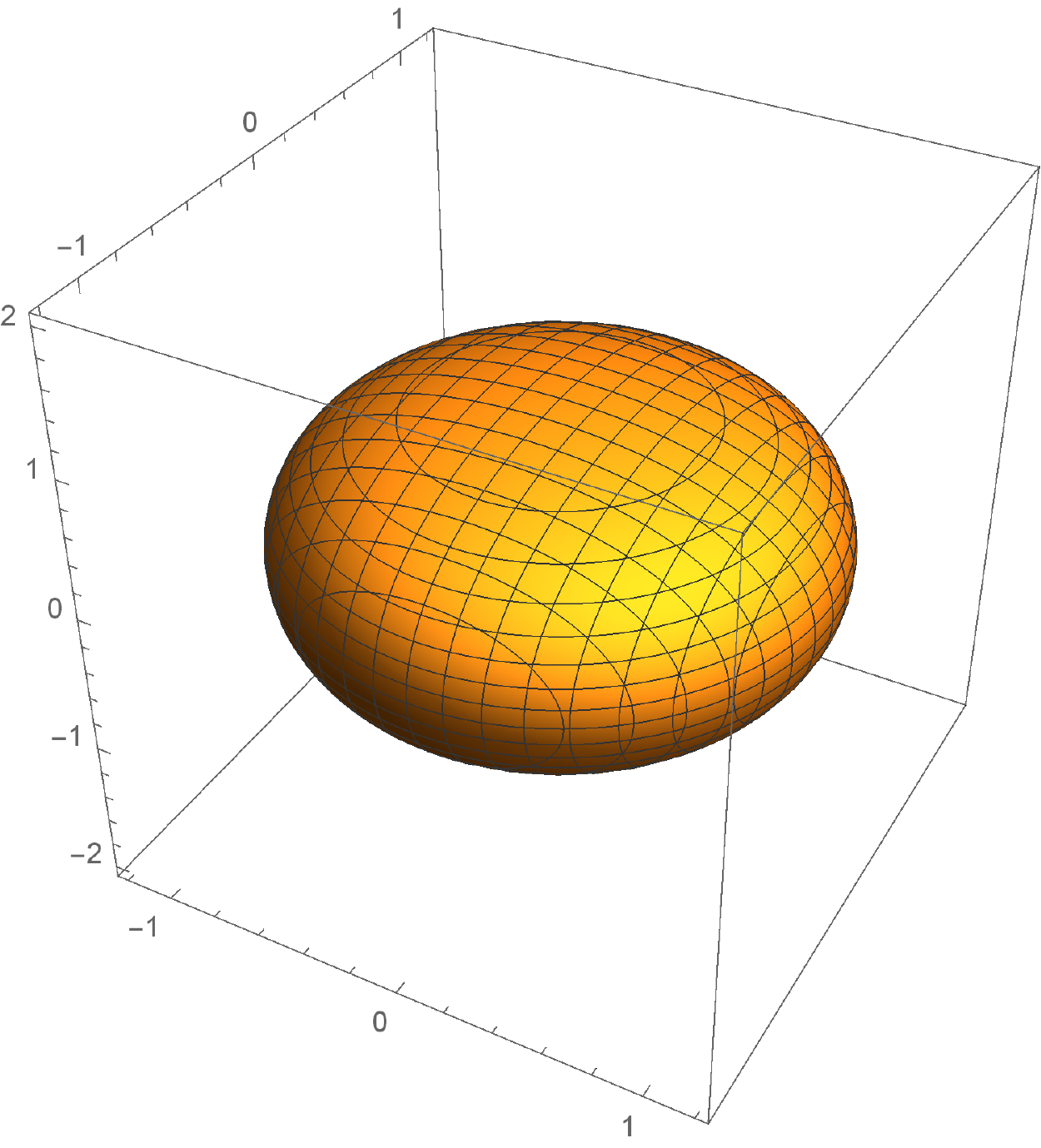}}\\
\subfigure[$q=e^{2}(\gamma=2)$]{
\label{Fig.sub.3}
\includegraphics[width=0.4\columnwidth]
{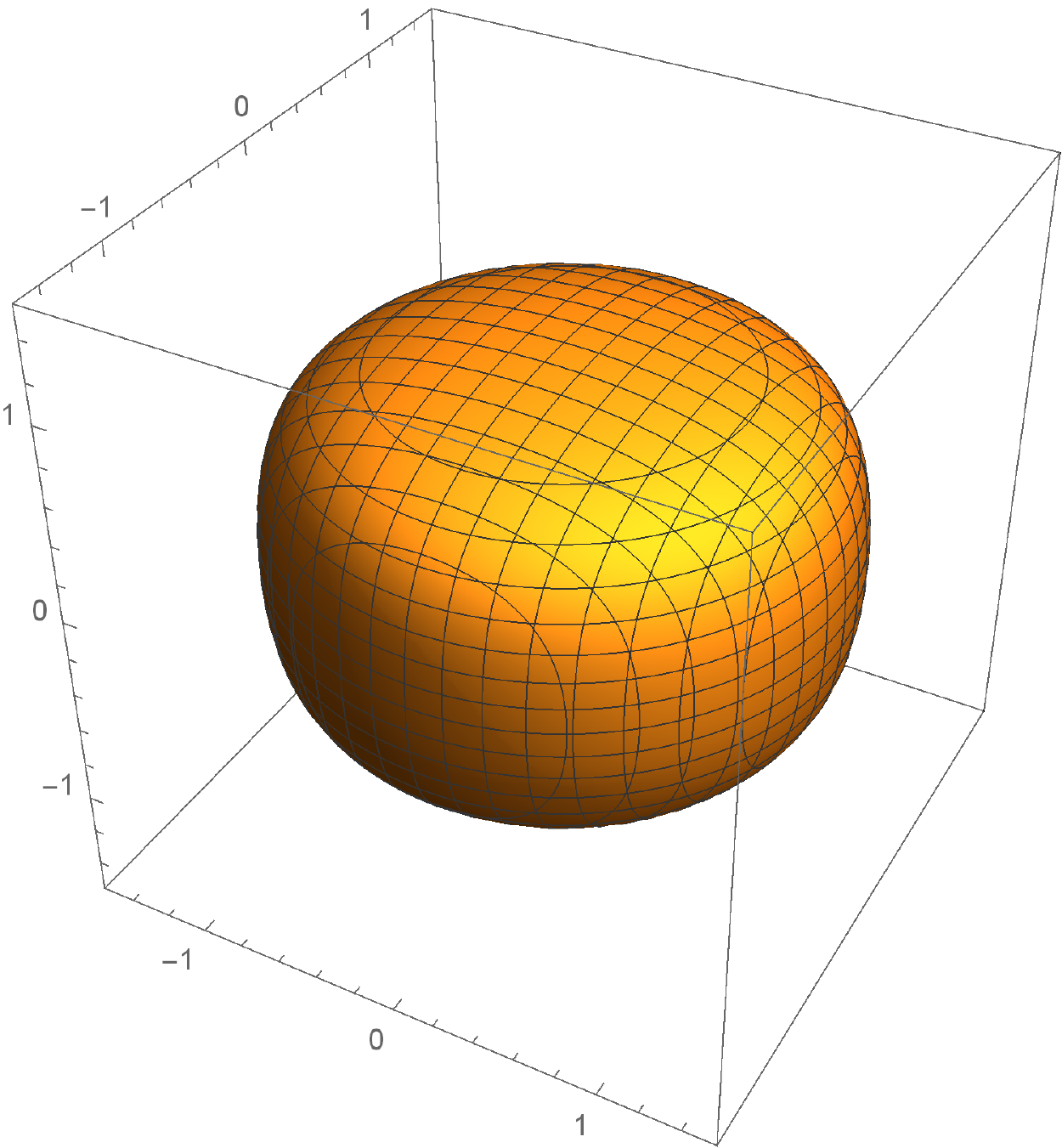}}
\subfigure[$q=e^{5}(\gamma=5)$]{
\label{Fig.sub.4}
\includegraphics[width=0.4\columnwidth]
{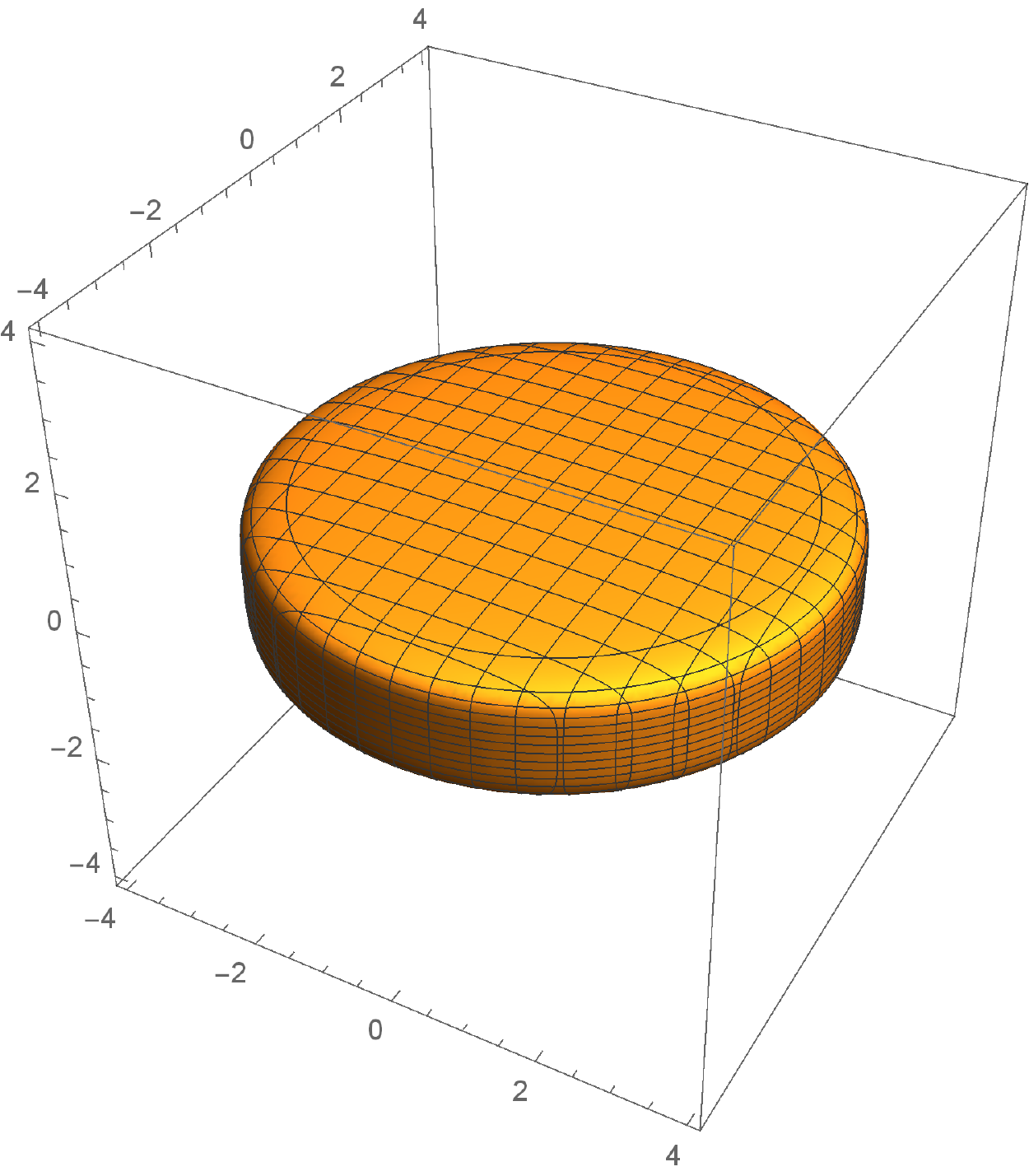}}
\caption{The $q$-deformation of the sphere: sphere Fig. \ref{Fig.sub.1} without deformation, ellipsoid like Fig. \ref{Fig.sub.2} for $\gamma=1$, drum like Fig. \ref{Fig.sub.3} for $\gamma=2$, cylinder like Fig. \ref{Fig.sub.4} for $\gamma=5$.}
\label{Fig.1}
\end{figure}

The Casimir operator $J$ of $\mathfrak{su}(2)$ is given by
\be\label{s3}
J^2=J^2_1+J^2_2+J^2_3.
\ee
The q-deformed Casimir operator $J'$ with respect to $\mathfrak{su}_q(2)$ has the form,
\be\label{qs3}
J'^{2}=J'^2_1+J'^2_2+\frac{(\sinh\gamma J'_3)^2}{\gamma\sinh\gamma}.
\ee
Taking $J_i$ ($J'_i$), $i=1,2,3$, as three real variables, and the Casimir operator $J^2$ ($J'^2$)
as a constant, the equation (\ref{s3}) ((\ref{qs3})) can be viewed as a three dimensional sphere (deformed sphere).
In fact, from symplectic geometry it has been shown in \cite{Fei} that $J_i$ ($J'_i$), $i=1,2,3$, constitute exactly the Poisson algebra of $\mathfrak{su}(2)$  ($\mathfrak{su}_q(2)$), which give rise further to the algebra $\mathfrak{su}(2)$  ($\mathfrak{su}_q(2)$) by geometrical quantization.
Therefore, hidden in the symmetry deformation is the deformation from a sphere (\ref{s3})
to deformed sphere (\ref{qs3}) which could be ellipsoid like, drum like, or even cylinder like depending on the deformation parameter $q=e^\gamma$, see Fig. \ref{Fig.1}.

\begin{figure}[!htb]
  \centering
  \includegraphics[width=0.6\columnwidth]
  {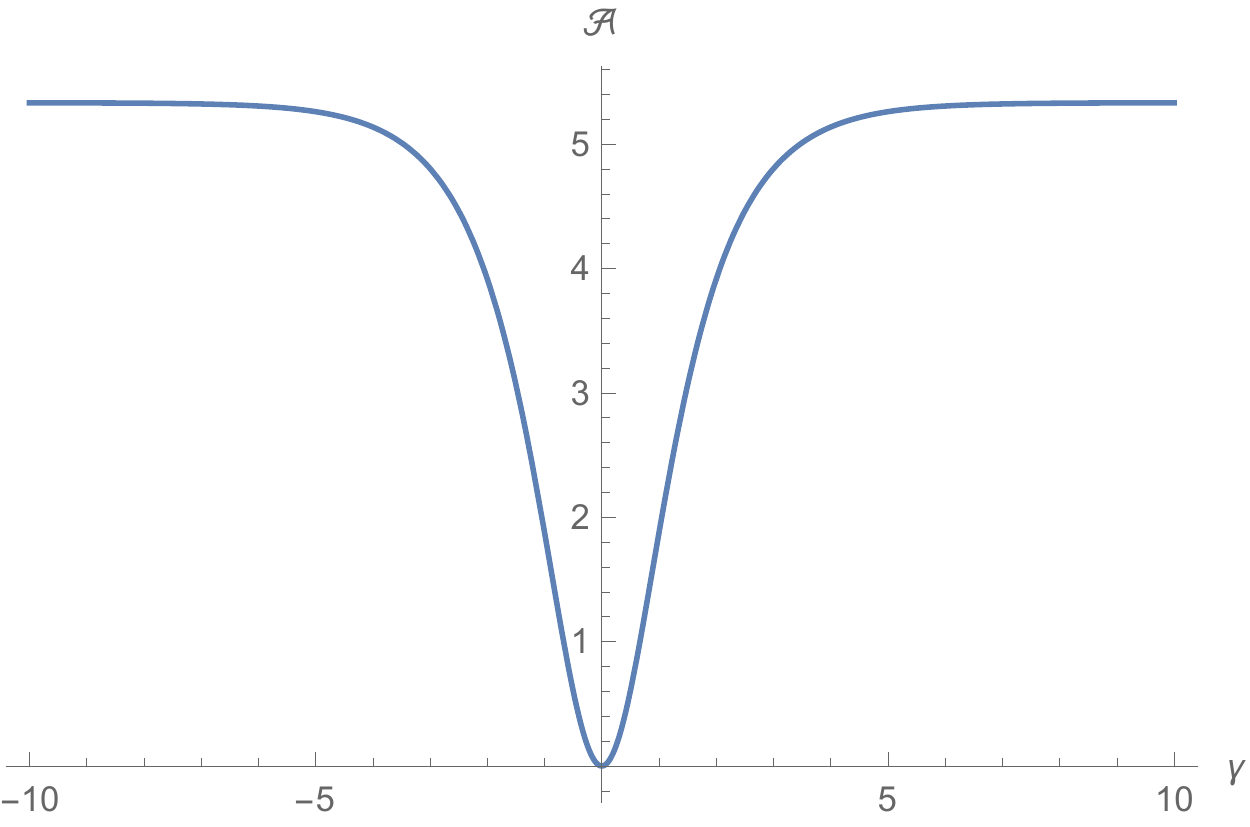}\\
  \caption{The asymmetry degree of the operator $J'^2$ with respect to the algebra $\mathfrak{su}(2)$, as a function of the deformation parameter $\gamma$.}\label{fig2}
\end{figure}

An interesting question is what the asymmetry degree of such ellipsoid like, drum like, or cylinder like geometries with respect to a perfect sphere is.
This kind of asymmetries can be characterized by asymmetries of $\mathfrak{su}_q(2)$ with respect to $\mathfrak{su}(2)$. To calculate the asymmetry of the operator $J'^2$ with respect to the algebra
$\mathfrak{su}(2)$, we consider the co-product representations \cite{Chaichian,Lie} of $\mathfrak{su}(2)$
\be\label{copsu2}
\begin{aligned}
&J_{\pm}=\sigma_{\pm}\otimes I+I\otimes\sigma_{\pm},\\
&J_3=\sigma_z\otimes I+I\otimes\sigma_z,
\end{aligned}
\ee
and $\mathfrak{su}_q(2)$
\be\label{copsuq2}
\begin{aligned}
&J'_{\pm}=\sigma_{\pm}\otimes q^{\sigma_z}+q^{-\sigma_z}\otimes
\sigma_{\pm},\\
&J'_3=J_3=\sigma_z\otimes I+I\otimes\sigma_z,
\end{aligned}
\ee
where  $\sigma_x=\frac{1}{2}
\begin{pmatrix}
0&1\\1&0
\end{pmatrix}$,
$\sigma_y=\frac{1}{2}
\begin{pmatrix}
0&-i\\i&0
\end{pmatrix}$,
$\sigma_z=\frac{1}{2}
\begin{pmatrix}
1&0\\0&-1
\end{pmatrix}$ are Pauli matrices, $I$ is the identity and $\sigma_{\pm}=\sigma_x\pm i\sigma_y$. With the co-product representations (\ref{copsu2}) and (\ref{copsuq2}) we have
\be\nonumber
J'^2=
\begin{pmatrix}
[\frac{3}{2}]^2_q&0&0&0\\
0&e^{\gamma}+[\frac{1}{2}]^2_q&1&0\\
0&1&e^{-\gamma}+[\frac{1}{2}]^2_q&0\\
0&0&0&[\frac{3}{2}]^2_q
\end{pmatrix},
\ee
where $[x]_q=\frac{x^q-x^{-q}}{q-q^{-1}}$ and $\gamma=\log q$,
\be
\begin{aligned}
&[J'^2,J_+]=
\begin{pmatrix}
0&e^{\gamma}-1&e^{-\gamma}-1&0\\
0&0&0&1-e^{\gamma}\\
0&0&0&1-e^{-\gamma}\\
0&0&0&0
\end{pmatrix},\\
&[J'^2,J_-]=
\begin{pmatrix}
0&0&0&0\\
1-e^{\gamma}&0&0&0\\
1-e^{-\gamma}&0&0&0\\
0&e^{\gamma}-1&e^{-\gamma}-1&0
\end{pmatrix},\\
&[J'^2,J_3]=0.
\end{aligned}
\ee
Then the asymmetry degree of $J'^2$ with respect to $\mathfrak{su}(2)$ is derived as
\be
\begin{aligned}
\mathcal{A}(\mathfrak{su}(2),J'^2)
&=\frac{1}{\|\tilde{J'^2}\|^2}
\sum_{\alpha=\pm,3}\|[J',J_{\alpha}]\|^2\\
&=\frac{16(\cosh\gamma-1)}{3\cosh\gamma}.
\end{aligned}
\ee
In Fig. \ref{fig2} we show the asymmetry degree of $J'^2$ with respect to $\mathfrak{su}(2)$ as a function of the deformation parameter $\gamma$. One can see that under the deformation
$\mathcal{A}(\mathfrak{su}(2),J'^2)$ increases quickly and finally tends to be stable as $|\gamma|$ increases.

\subsection{The integrable spin chain with boundary conditions}

The Heisenberg spin chain models play important roles in the study of critical points and phase transitions of magnetic systems.
The Hamiltonian of the standard $XXX$ $1/2$-spin chain model with nearest-neighbor interactions
has the form
\be\label{xxx}
H_{XXX}=\sum^N_{j=1}(\sigma^x_j \sigma^x_{j+1}+\sigma^y_j\sigma^y_{j+1}+
\sigma^z_j\sigma^z_{j+1}).
\ee
Generalizing the co-product representation (\ref{copsu2}) of $\mathfrak{su}(2)$ to $N$-spin case, one obtains
\be\label{su2t}
J_{\pm}=\frac{1}{2}\sum^N_{j=1}\sigma^{\pm}_j,
~~~~J_3=\frac{1}{2}\sum^N_{j=1}\sigma^z_j,
\ee
where the operator $\sigma^{\alpha}_j$ means acting $\sigma^{\alpha}$ on the space of the $j$th spin, $\sigma^{\alpha}_j=I\otimes I...\otimes \sigma^{\alpha}_j\otimes ...\otimes I~(\alpha=\pm,3)$. It is easily verified that $H_{XXX}$ is $\mathfrak{su}(2)$-symmetric, $[H_{XXX},J_{\pm}]=[H_{XXX},J_{3}]=0$.

Nevertheless, the following integrable spin chain Hamiltonian $H_q$ with an extra boundary term is no longer $\mathfrak{su}(2)$-symmetric,
\be\label{hq}
\begin{aligned}
H_q=\sum^N_{j=1}&(\sigma^x_j
\sigma^x_{j+1}+
\sigma^y_j\sigma^y_{j+1}+
\frac{q+q^{-1}}{2}\sigma^z_j
\sigma^z_{j+1})\\
&\quad
+\frac{q-q^{-1}}{2}(\sigma^z_1-
\sigma^z_N).
\end{aligned}
\ee
This Hamiltonian, derived from the solution of Yang-Baxter equations \cite{Kulish1}, reduces to $H_{XXX}$ when $q$ goes to $1$.

Generally, $H_q$ is symmetric with respect to the $q$-deformed algebra $\mathfrak{su}_q(2)$.
From the co-product representation (\ref{copsuq2}) of $\mathfrak{su}_q(2)$, for the $N$-spin case one has the generators of $\mathfrak{su}_q(2)$,
$$
J'_{\pm}=\frac{1}{2}\sum^N_{j=1}\sigma^{'\pm}_j,
~~~~J'_3=J_3=\frac{1}{2}\sum^N_{j=1}\sigma^z_j,
$$
where $\sigma'_j=q^{-\sigma_z}\otimes \ldots\otimes q^{-\sigma_z}\otimes \sigma_j\otimes q^{\sigma_z}\otimes ...\otimes q^{\sigma_z}$. It is easily verified that $H_{q}$ is $\mathfrak{su}_q(2)$-symmetric, $[H_{q},J'_{\pm}]=[H_{q},J'_{3}]=0$.

According to the representation (\ref{su2t}) of $\mathfrak{su}(2)$, we get
\begin{widetext}
\be
\begin{aligned}
&[H_q,J_+]=\frac{1}{2}\Big[(\cosh\gamma-1)
\sum^N_{k=1}(\sigma^z_k\sigma^+_{k+1}+
\sigma^+_k\sigma^z_{k+1})+
\sinh\gamma(\sigma^+_1-\sigma^+_N)\Big],\\
&[H_q,J_-]=\frac{-1}{2}\Big[(\cosh\gamma-1)
\sum^N_{k=1}(\sigma^z_k\sigma^-_{k+1}+
\sigma^-_k\sigma^z_{k+1})+
\sinh\gamma(\sigma^-_1-\sigma^-_N)\Big],\\
&[H_q,J_3]=0.
\end{aligned}
\ee
\end{widetext}

Then the asymmetry of $H_q$ with respect to $\mathfrak{su}(2)$ can be obtained,
\be\label{asymmetryYB}
\begin{aligned}
\mathcal{A}&(\mathfrak{su}(2),H_q)=
\frac{1}{\|\tilde{H_q}\|^2}
\sum_{\beta=\pm,3}\|[H_q,J_{\beta}]\|^2\\
&=\frac{N(\cosh\gamma-1)^2+4\sinh^2\gamma}
{N(2+\cosh^2\gamma)
+8\sinh^2\gamma}.
\end{aligned}
\ee

\begin{figure}[htb]
  \centering
  \includegraphics[width=0.9\columnwidth,
  height=5cm]{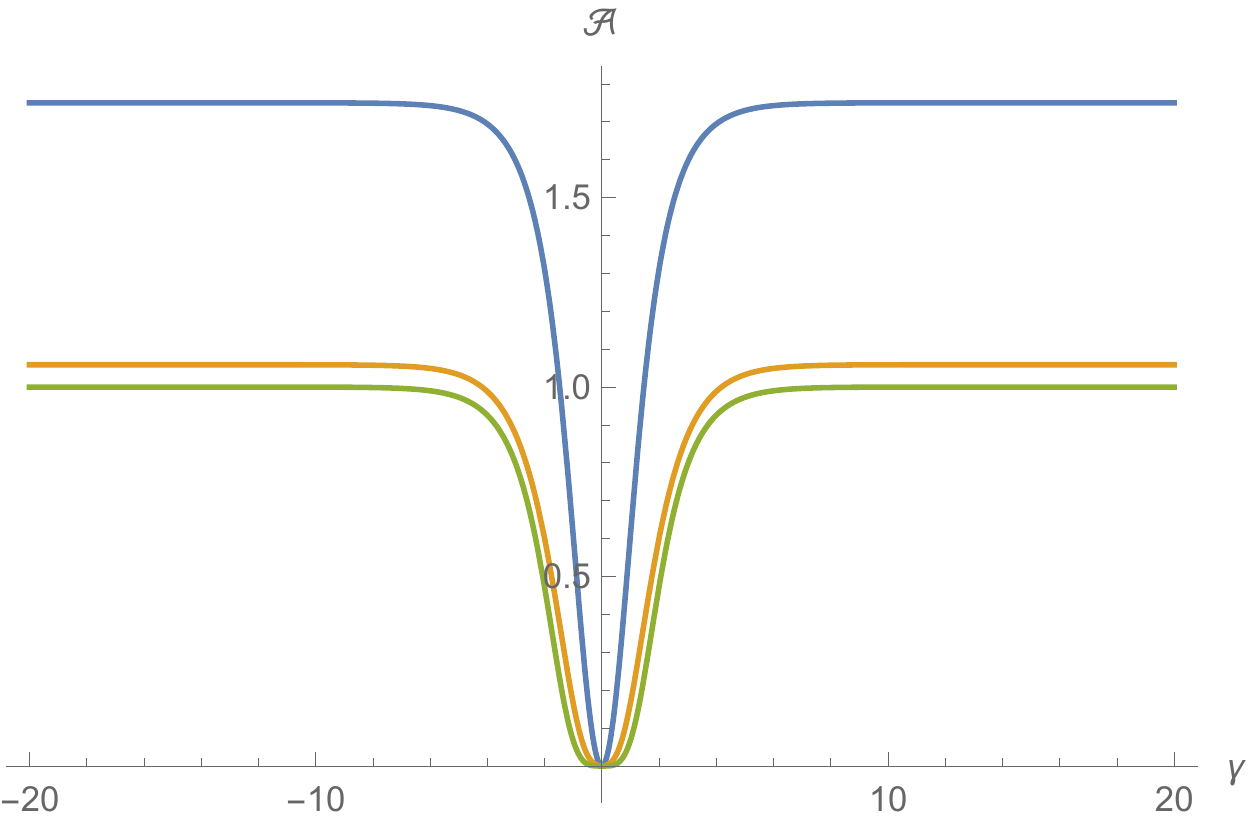}\\
  \caption{The $\mathcal{A}(\mathfrak{su}(2),H_q)$ with $N=3$~(blue curve), $N=50$~(orange curve), and $N\rightarrow\infty$~(green curve) as the functions of the deformation parameter $\gamma$.}\label{fig3}
\end{figure}

From the formula (\ref{asymmetryYB}) we have that the asymmetry degree of $H_q$ with respect to $\mathfrak{su}(2)$ depends on the number of spines as well. In particular, when $N\rightarrow \infty$, we have
$$
\mathcal{A}(\mathfrak{su}(2),H_q)
\rightarrow\frac{(\cosh\gamma-1)^2}
{\cosh^2\gamma+2}.
$$
The asymmetry $\mathcal{A}(\mathfrak{su}(2),H_q)$ vs the deformation parameter $\gamma$ is shown in Fig. \ref{fig3} for different spin number $N$.
From Fig. \ref{fig3} one sees that the asymmetry of $H_q$ with respect to $\mathfrak{su}(2)$
varies with $\gamma$ in the way similar to that of $J'^2$. As $|\gamma|$ increases the asymmetry degree increases sharply, and eventually tends to be stable. In addition from Fig. \ref{fig2} and Fig. \ref{fig3} we also find that the symmetry of $J'^2$ and $H_q$ with respect to $\mathfrak{su}(2)$ are broken symmetrically with respect to deformation parameter $\gamma$.

\section{Discussions and Conclusions}

In \cite{Fang} the asymmetry degree $A(G,H)$ of a Hamiltonian $H$ with respect to a Lie group $G$ has been defined in terms of the commutators of the Hamiltonian and the group elements.
Let $\mathfrak{g}$ be the Lie algebra associated with the Lie group $G$.
In an infinitesimal neighborhood of the unit element $e$ of $G$, one can show that
$A(G,H)$ is proportional to $\mathcal{A}(\mathfrak{g},H)$ defined in (\ref{def1}).
Nevertheless, the asymmetry $\mathcal{A}(\mathfrak{g},H)$ of an operator $H$ with respect to an algebra $\mathfrak{g}$ defined in (\ref{def1}) is for any algebras $\mathfrak{g}$ including Lie algebras and quantum algebras etc.. The relation between $A(G,H)$ and $\mathcal{A}(\mathfrak{g},H)$ for general $\mathfrak{g}$ could be more complicated.

We have studied the asymmetry degree of operators with respect to certain algebraic representations. The asymmetry has been defined and its properties have been investigated. As detailed applications, the asymmetries of the q-deformed harmonic Hamiltonian system (\ref{hp}) with respect to the harmonic operator representation (\ref{hsu2}) of $\mathfrak{su}(2)$, the q-deformed Casimir operator (\ref{qs3}) with respect to the co-product representation (\ref{copsu2}) of $\mathfrak{su}(2)$,
and the integrable spin chain Hamiltonian (\ref{hq}) with respect to the $\mathfrak{su}(2)$ representation (\ref{su2t}) have been computed analytically. In particular, the asymmetry
(\ref{def1}) presents a qualitative characterization of the degree of geometrical deformation of sphere. As symmetry plays crucial roles in physics, our results may highlight investigations on some important topics such as critical phenomena and phase transitions in physical systems.
In addition, the ${\cal PT}$-symmetric quantum mechanics has been extensively studied recently \cite{Ruschhaupt,SimonMath,Simonpra}.
In such pseudo-Hermitian quantum mechanical systems,
the Hamiltonian are no longer necessarily Hermitian, but may still
have real eigenvalues \cite{14,15,16}.
Despite the original motivation to build a new framework
of quantum theory, researchers are also aware of the
importance of simulating the ${\cal PT}$-symmetric systems with conventional quantum mechanics \cite{hmy}.
Adapting our proposed asymmetry degree to non-Hermitian Hamiltonians would be also an interesting open question.

\medskip
\noindent{\bf Acknowledgments}\, \,
This work is supported by the NSFC (No. 11701128, No. 11847244, No. 11675113), Beijing Municipal Commission of Education (KZ201810028042), and Beijing Natural Science Foundation (Z190005).


\begin{thebibliography}{9}

\bibitem{Wigner}E. P. Wigner, Ann. Math. {\bf40}, 149 (1939).

\bibitem{Wigners}E. P. Wigner, \emph{Symmetries and Reflections}, University Press (1965);\\
 E. P. Wigner, \emph{Group Theory and Its Application to the Quantum Mechanics of Atomic Spectra}, Cambridge University Press (1959).


\bibitem{Dirac}P. A. M. Dirac, \emph{The Principles of Quantum Mechanics}, 3rd. edition, Oxford University Press (1958)

\bibitem{Noether} E. N\"{o}ther, \emph{Invariante Variationsprobleme} (1918) in German;  Mort Tavel (translator), \emph{Invariant Variation Problems}, Transport Theory and Statistical Physics. {\bf1} (3): 186¨C207 (1971).

\bibitem{Weinberg} S. Weinberg, \emph{Lectures on Quantum Mechanics}, Cambridge University Press (2013).

\bibitem{Landau} L. Landau and E. M. Lifsshitz, \emph{Quantum Mechanics. Nonrelativistic Theory} 3rd edition (translated from Russian by J. B. Sykes and J. S. Bell) Pergamin Press Ltd, U. K. (1977).

\bibitem{Weyl} H. Weyl, \emph{The Classical Groups}, Princeton University Press, New Jersey (1946).

\bibitem{Hamermesh} Hamermesh, \emph{Group Theory and Its Application to Physical Problems}, Addison wesley publishing company, U. S. A. (1962).

\bibitem{Kulish} P. P. Kulish and N. Yu. Reshetikhm, J. Soviet Math. {\bf23}, 2435 (1983). (translation from: \emph{Zapiski Nauch. Seminarov LOMI} {\bf101} (1981).)

\bibitem{Faddeev} L. D. Faddeev and L. A. Takhtajan, \emph{Lecture Notes in Physics} {\bf 246} (1986) 166.

\bibitem{Chaichian} M. Chaichian, A. Demichev, \emph{Introduction to Quantum Groups} World Scientific (1996).

\bibitem{Macfarlane} J. A. Macfarlane, J. Phys. A: Math. Gen. {\bf22} (1989) 4581-4588.

\bibitem{Biedenharn} L. C. Biedenharn, J. Phys. A: Math. Gen. {\bf22} (1989) L873-L878.

\bibitem{Sun} C. P. Sun and H. C. Fu, J. Phys. A: Math. Gen. {\bf22} (1989) L983-L986.

%\bibitem{Lie} \emph{Introduction to Lie Algebras and its Representation}

%\bibitem{Penrose} O. Penrose and L. Onsager, Phys. Rev. {\bf 104} 576 (1956).
%\bibitem{Leggett} A. J. Leggett and F. Sols, Found. Phys. {\bf 21} 353 (1991).
%\bibitem{Nambu} Y. Nambu and G. Jona-Lasinio, Phys. Rev. {\bf 122} 345 (1961).

\bibitem{Gour} G. Gour and R. W. Spekkens, New J. Phys. {\bf10}, 033023 (2007);\\
 G. Gour, I. Marvian, and R. W. Spekkens, Phys. Rev. A {\bf 80}, 012307 (2009).


\bibitem{Vaccaro} J. A. Vaccaro, F. Anselmi, H. M. Wiseman and K. Jacobs, Phys. Rev. A {\bf 77}, 032114 (2008).

\bibitem{Toloui} B. Toloui, G. Gour, and B. C. Sanders, Phys. Rev. A {\bf84}, 022322 (2011).

\bibitem{Skotiniotis} M. Skotiniotis and G. Gour, New J. Phys. {\bf14}, 073022 (2012).

\bibitem{Iman} I. Marvian and R. W. Spekkens, New J. Phys. {\bf 15}, 033001 (2013);\\
I. Marvian and R. W. Spekkens, Phys. Rev. A {\bf 90}, 014102 (2014);\\
I. Marvian and R. W. Spekkens, Nat. Commun. {\bf 5}, 3821 (2014);\\
I. Marvian and R. W. Spekkens, arXiv:1212. 3378 (2012).

\bibitem{Piani} M. Piani, M. Cianciaruso, T. R. Bromley, C. Napoli, N. Johnston, and G. Adesso, Phys. Rev. A {\bf 93}, 042107 (2016).

\bibitem{Luo} S. L. Luo and Y. Sun, Phys. Rev. A {\bf98}, 012113 (2018).

\bibitem{Wakakuwa} E. Wakakuwa, Phys. Rev. A {\bf 95} 032328 (2017).

\bibitem{Fang} Y. N. Fang, G. H. Dong, D. L. Zhou, and C. P. Sun, Common. Theor. Phys. {\bf 65} (2016) 423-433.

\bibitem{Dong} H. G. Dong, Y. N. Fang, C. P. Sun, Commun. Theor. Phys, \textbf{68} (2017) 405-411.

\bibitem{Dong2} H. G. Dong, Z. W. Zhang, C. P. Sun and Z. R. Gong, Sci. Rep. {\bf 7}, 12947 (2017).

\bibitem{Hopf} E. Abe, \emph{Hopf Algebra} Cambridge Tracts in Math., No. 74, Cmbridge Unvi. Press, Cambridge-New York (1980).

\bibitem{Heisenberg} W. Heisenberg (1954), as quoted in H. P. D\"{u}rr, \emph{Werner Heinsenberg und die Physik unserer Zeit}, (S. 299, Fr. Vieweg u. Sohn, Braunschweig, 1961.)

\bibitem{Schmidke} W. B. Schmidke, J. Wess and B. Zumino, Z. Phys. C -Particles and Fields {\bf 52}, 471-476 (1991).

\bibitem{Jimbo1} M. Jimbo, Lett. Math. Phys. {\bf10} 63 (1985); Jimbo M, Commun. Math. Phys. {\bf102} 537 (1986);

\bibitem{Drinfeld}  V. G. Drinfeld, Proc. Int. Conf. Math., Berkeley, p798;

\bibitem{Chang} Z. Chang, W. Chen and H. Y. Guo, J. Phys. A: Math. Gen. {\bf 23} 4185-4190 (1990);\\ Z. Chang, W. Chen, H. Y. Guo and H. Yan, J. Phys. A: Math. Gen. {\bf 23} 5371-5382 (1990);\\
    Z. Chang, S. M. Fei, H. Y. Guo and H. Yan, J. Phys. A: Math. Gen. {\bf 24} 5435-5444 (1991).

\bibitem{Fei} S. M. Fei and H. Y. Guo, J. Phys. A: Math. Gen. {\bf 24} (1991) 1-10;\\
 S. M. Fei, J. Phys. A: Math. Gen. {\bf 24} (1991) 5195-5214.

\bibitem{Lie} J. E. Humphreys, \emph{Introduction to Lie Algebras and Representation Theory}, Sringer-Verlag New York Inc. (1972).

\bibitem{Kulish1} P. P. Kulish, N. Yu. Reshetikhm, and E. K. Sklyanin, Lett. Math. Phys. {\bf5}, 393 (1981);\\
    P. P. Kulish and E. K. Sklyanin, J. Soviet Math. {\bf19}, 1596 (1982).

\bibitem{Ruschhaupt} A. Ruschhaupt, T. Dowdall, M. A. Sim\'{o}n and J. G. Muga, Europhysics Letters {\bf 120} 20001 (2017).
    
\bibitem{SimonMath} M. A. Sim\'{o}n, A. Buend\'{i}a and J. G. Muga, Mathematics {\bf6} 111 (2018).
    
\bibitem{Simonpra} M. A. Sim\'{o}n, A. Buend\'{i}a, A. Kiely, Ali Mostafazadeh, and J. G. Muga,  Phys. Rev. A {\bf99}, 052110 (2019).

\bibitem{14}L. Ge, A. D. Stone, Phys. Rev. X. {\bf4}, 031011 (2014).

\bibitem{15}T. Theurer, N. Killoran, D. Egloff, and M. B. Plenio, Phys. Rev. Lett. {\bf119}, 230401 (2017).

\bibitem{16}B. Qi, L. Zhang, L. Ge, Phys. Rev. Lett. {\bf120}, 093901 (2018).

\bibitem{hmy} M. Y. Huang, R. K. Lee, L. J. Zhang, S. M. Fei, J. D. Wu, Phys. Rev. Lett. {\bf123}, 080404 (2019).

\end{thebibliography}
\end{document}